\documentclass[aps,prl,twocolumn,english,showpacs,10pt]{revtex4-1}
\usepackage{amsmath,bm}
\usepackage{amssymb}
\usepackage{graphicx}
\usepackage{babel}
\usepackage{cases}
\usepackage{natbib}
\usepackage{floatrow}

\makeatletter

\begin{document}

\title{Theory of Subradiant States of a One-Dimensional Two-Level Atom Chain}
\author{Yu-Xiang Zhang}
\email{iyxz@phys.au.dk}
\author{Klaus M{\o}lmer}
\email{moelmer@phys.au.dk}
\affiliation{Department of Physics and Astronomy, Aarhus University, 8000 Aarhus C, Denmark}
\date{\today}

\begin{abstract}
Recently, the subradiant states of one-dimensional two-level atom chains coupled to light modes were found to
have decay rates obeying a universal scaling, and an unexpected fermionic character of the multiply-excited
subradiant states was discovered. In this Letter, we theoretically obtain the singly-excited subradiant states, and by
eliminating the superradiant modes, we demonstrate a relation between the multiply-excited subradiant states and the Tonks-Girardeau
limit of the Lieb-Liniger model which explains the fermionic behavior. In addition, we identify a new family of states with correlations different from the fermionic ansatz.
\end{abstract}
\maketitle

To achieve controllable and deterministic photon-atom interfaces for applications in quantum information
processing and quantum sensing, large atom ensembles may be used to enhance the coupling to photons \cite{Hammerer2010}.
The photons induce both coherent and dissipative atom-atom interactions that
yield collective phenomena of super- or sub-radiance \cite{VanLoo2013}, wherein a collective excitation of
the atom ensemble decays faster or slower than individual atomic excitations. While superradiance has been extensively studied since
the seminal work of Dicke \cite{Dicke1954}, subradiance of a large ensemble was observed
only very recently in cold atom clouds \cite{Guerin2016,Weiss2018} and metamaterial arrays \cite{Jenkins2017}.
Comprehensive theoretical tools for the subradiance are still elusive \cite{Plankensteiner:2015aa,Asenjo-Garcia2017,Albrecht2018,Henriet:2019aa}
due to the complicated long-range interactions and many-body
features of the atomic ensembles \cite{Solano2017,Noh2016,Olmos:2013aa}.
A one-dimensional (1D) chain of equally spaced two-level atoms
offers the simplest geometry to gain insight in the collective decay mechanisms, and implementation of
such chains coupled to nanofibers \cite{Kornovan2016}, 1D waveguides \cite{Haakh2016,Ruostekoski2016,Ruostekoski:2017aa,Zoubi:2014aa},
and the full
vacuum electromagnetic field in 3D free space \cite{Sutherland2016,Olmos:2013aa,Bettles:2016aa,Jen:2016aa,Asenjo-Garcia2017} has attracted considerable attention.
Super- and subradiance phenomena are in these systems supplemented by further interesting properties and applications
such as atomic mirrors \cite{Chang2012}, photon Fock state synthesis \cite{Gonzalez-Tudela2017},
enhancement of cooperativity \cite{Plankensteiner2017} and applications in
quantum computation \cite{Paulisch2016}.

Recently, the subradiant states of such 1D chains of $N$ qubits in 3D
free space and coupled to 1D waveguide were numerically found to have
a series of seemingly universal properties \cite{Asenjo-Garcia2017,Albrecht2018,Henriet:2019aa}:
In the one-excitation sector where only one of the $N$ atoms is excited,
if we sort all eigenstates (to be elaborated) by increasing decay rates with integer labels from $\xi=1$ to $\xi=N$, the most
subradiant states ($\xi\ll N$) have decay rates $\gamma_{\xi}\propto \xi^2/N^3$.
In the multi-excitation sectors,
the most subradiant states have a \emph{fermionic} character, e.g.,
a most subradiant state with two excitations is given by
$|F_{1,2}\rangle\propto\sum_{i<j}(c_{1,i}c_{2,j}-c_{1,j}c_{2,i})|e_i, e_j\rangle$,
built from subradiant states $|\psi_{1(2)}\rangle=\sum_{i}c_{1(2),i}|e_i\rangle$ in the one-excitation sector, where $|e_i\rangle$ ($|e_i, e_j\rangle$) represents the state with the $i$th ($i$th and the $j$th) atom excited to $|e\rangle$ while all other atoms are in the ground state $|g\rangle$.
The decay rate of $|F_{1,2}\rangle$ is the sum of the decay rates of $|\psi_{1}\rangle$ and $|\psi_2\rangle$.
The infidelity of the fermionic ansatz $|F_{1,2}\rangle$ to exact numerical results
scale as $N^{-2}$ and $N^{-1}$ for two different classes of states \cite{Albrecht2018}.

\begin{figure}[b]
  \centering
    \includegraphics[width=0.95\textwidth]{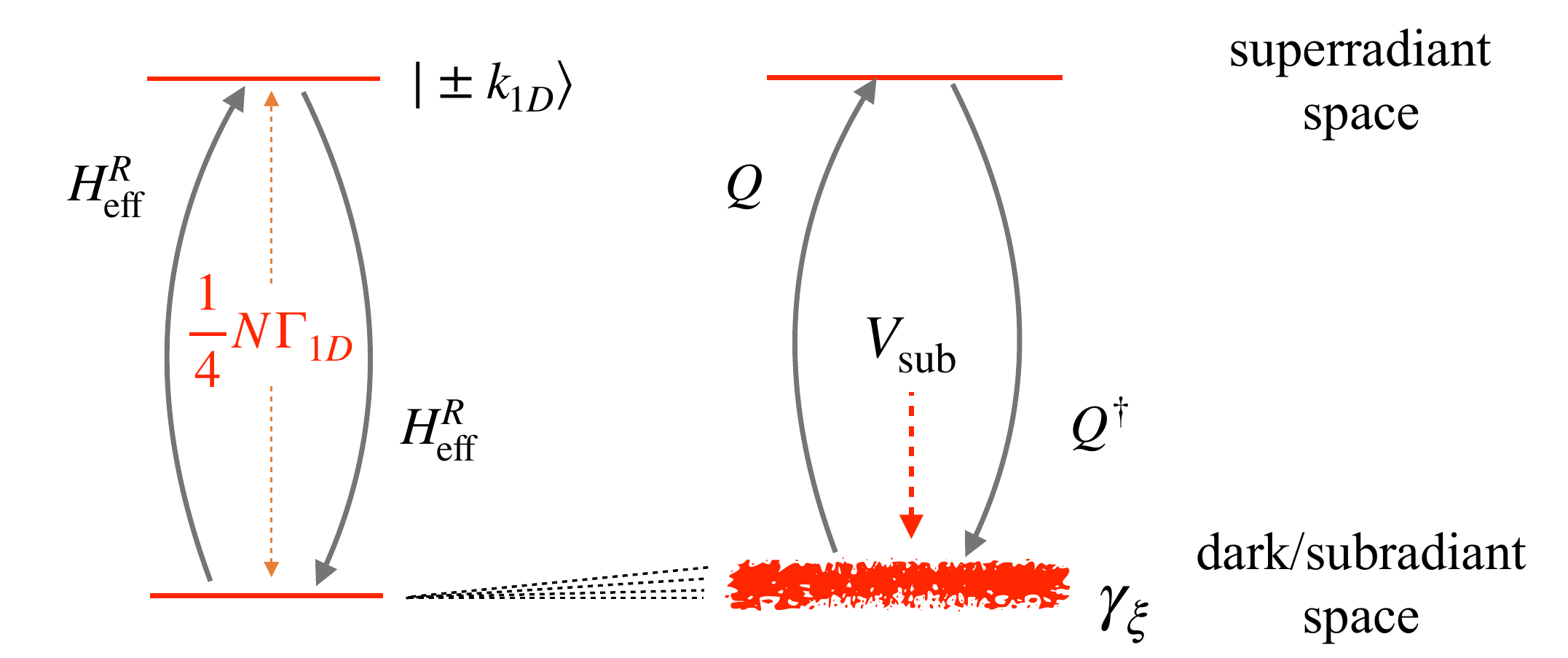}
\caption{Outline of the theory: The imaginary part of the effective atomic Hamiltonian, $H^{I}_{\mathrm{eff}}$,
identifies superradiant ($|\pm k_{1D}\rangle$) and dark sub spaces of states (left panel), coupled perturbatively by $H^{R}_{\mathrm{eff}}$
to produce the subradiant modes with decay rates $\gamma_{\xi}\propto \xi^2/N^3\,(\xi\ll N)$.
The Holstein-Primakoff transformation bosonizes the super- and subradiant modes and introduces a coupling ($Q$ and $Q^{\dagger}$) between them (right panel). The coupling
effectively yields a strong interaction $V_{\mathrm{sub}}$ among the multiply excited subradiant modes leading to the formation of states with fermionic character.}
\label{fig1}
\end{figure}

It is intriguing why these properties appear for both the infinite range atom-atom interactions
mediated by 1D guided fields \cite{Albrecht2018,Henriet:2019aa}, and the long-range ($\sim 1/r$) or short-range ($\sim 1/r^3$) interactions
mediated by the 3D free-space field \cite{Asenjo-Garcia2017,Henriet:2019aa}. A thorough theoretical understanding is needed to guide further experimental studies and applications of subradiance. In this Letter, we provide such understanding based on the theoretical treatment of the physics summarized in Fig. \ref{fig1}.

\emph{Spin Models}- For light-matter interactions
where the Markovian approximation is applicable, we can eliminate the light modes to obtain a master
equation describing only the atoms \cite{Dung2002}. The master equation is equivalent to a Monte Carlo
wave function formalism \cite{Moelmer1993}, where the atomic state evolves stochastically under quantum jumps and deterministically
under $H=H_{0}+ H_{\text{eff}}$,
where $H_0$ is the bare Hamiltonian of the atoms and the \emph{non-Hermitian} $H_{\text{eff}}$
describes both coherent and dissipative atom-atom interactions mediated by the vacuum field.
The right eigenstates of $H$, or equivalently of $H_{\text{eff}}$ \cite{Asenjo-Garcia2017},
in each manifold of states with any given number of atomic excitations have
decay rates that are twice the imaginary part of the
corresponding right eigenvalues.
We focus our analysis on the qubit chain coupled to a 1D waveguide,
but our treatment provides sufficient insight to also account for the case of coupling to the 3D vacuum field.
For an atom chain coupled to a 1D waveguide,  we have \cite{Chang2012,Albrecht2018}:
\begin{equation}\label{H-1d}
H_{\text{eff}}=-\frac{i}{2}\Gamma_{1D}\sum_{m,n=1}^{N} e^{i k_{1D}|z_m-z_n|}\sigma_m^\dagger\sigma_n,
\end{equation}
where $\Gamma_{1D}$ is the decay rate of a single atom coupled to the waveguide \cite{Chang2012},
$k_{1D}$ is the wavenumber of the waveguide mode resonant with the atomic transition, and
$\sigma_m=|g\rangle_m\langle e|$ acts on the $m$th atom. We assume the
atoms are equidistantly spaced by $d$.
For convenience, we shall denote $H_{\mathrm{eff}} = H^{R}_{\text{eff}}-i H^{I}_{\text{eff}}$.

\emph{One-Excitation Sector}-The eigenstates of $H^{I}_{\text{eff}}$ divide the one-excitation sector into,
a two-dimensional superradiant subspace
spanned by Bloch states $|{\pm} k_{1D}\rangle
= N^{-1/2}\sum_{m=1}^{N}e^{\pm i k_{1D} z_m}|e_m\rangle$ with eigenvalue $N\Gamma_{1D}/4$;
and an $(N{-}2)$-dimensional dark space with eigenvalue 0, see Fig. \ref{fig1}.
The dark states acquire weak (subradiant) decay rates because of their admixture of superradiant states induced by
the perturbation from $H^{R}_{\text{eff}}$.

While the perturbation view is informative, a more direct approach to the subradiant states applies the following exact result for the
Bloch states $|k\rangle$ ($k\neq\pm k_D$),
\begin{equation}\label{Bloch-check-1}
H_{\text{eff}}\,|k\rangle=\omega_{k}|k\rangle- i\frac{\Gamma_{1D}}{2} (g_{k}|k_{1D}\rangle
-h_{k}|{-}k_{1D}\rangle),
\end{equation}
where $\omega_k{=}\frac{\Gamma_{1D}}{4}\sum_{\epsilon=\pm}\cot(\frac{k_{1D}+\epsilon k}{2}d)$, and the ``tails''
$g_{k}=\frac{e^{i(k-k_{1D})z_1}}{1-e^{i(k-k_{1D})d}}$ and $h_{k}=\frac{e^{i(k+k_{1D})z_N}}{e^{-i(k+k_{1D})d}-1}$.
It follows that a superposition of two degenerate states, $|k\rangle$ and $|{-}k\rangle$, is an eigenstate of $H_{\mathrm{eff}}$
with eigenvalue $\omega_k$ and has no tails if $k$ is a solution to the equation $g_{k}h_{-k}=g_{-k}h_{k}$.
This equation has only solutions for complex values of $k$. In the regimes $k\approx 0$ or $\pm \pi/d$ (center or edges of the first Brillouin zone), supposing
$k_{\xi}=0+\delta_{\xi}$ and $k_{\xi}=-\pi/d+\delta_{\xi}$ respectively, we find to order
$N^{-2}$,
\begin{equation}\label{k-complex}
\delta_{\xi}=\frac{\xi\pi}{Nd}\times
\begin{cases}
 1-i\frac{1}{N}\cot(\frac{k_{1D}}{2}d), & k\approx 0 \\
 1+i\frac{1}{N}\tan(\frac{k_{1D}}{2}d), & k\approx -\pi/d
\end{cases}
\end{equation}
with $\xi=1,2,3\cdots$, $\xi\ll N$. Note that Eq. (\ref{k-complex}) amounts to an $1/N^2$-order
imaginary correction to the Bloch wavenumber.

Next, we substitute Eq. (\ref{k-complex}) into the expression for $\omega_k$, which is parabolic near $k\approx 0$ and $\pm\pi/d$, i.e.,
$\omega_k\propto \delta_{\xi}^2$. Then the imaginary corrections directly
yield the $\xi^2/N^3$-scaling of the decay rates \cite{Albrecht2018}:
\begin{equation}\label{decay 1}
\gamma_{\xi}=\Gamma_{1D}\frac{\pi^2}{2}\frac{\xi^2}{N^3}\times
\begin{cases}
\frac{\cos^2(k_{1D} d/2)}{\sin^4(k_{1D} d/2)}, & k\approx 0\\
\frac{\sin^2(k_{1D} d/2)}{\cos^4(k_{1D} d/2)}, & k\approx -\pi/d.
\end{cases}
\end{equation}
The eigenstates are written as
\begin{equation}\label{ansatz}
\begin{aligned}
|\phi_{k_\xi}\rangle \propto & \; g_{-k_{\xi}}|k_{\xi}\rangle-g_{k_{\xi}}|{-}k_{\xi}\rangle \\
= & \frac{1}{\sqrt{2}}(\,|k^{(0)}_{\xi}\rangle-|{-}k^{(0)}_{\xi}\rangle\,) +O(\frac{\xi}{N}),
\end{aligned}
\end{equation}
where $k_{\xi}^{(0)}=\xi\pi/(Nd)$ or $-\pi/d+\xi\pi/(Nd)$.

\emph{Universality}-The $\xi^2/N^3$-scaling has also been numerically found for 1D atom chains coupled to
3D free-space modes \cite{Asenjo-Garcia2017,Albrecht2018,Henriet:2019aa}, where the effective Hamiltonian $H_{3D,\mathrm{eff}}$ is determined
by the Green's dyadic tensor (see the Supplemental Material). Fourier transformation of the Green's tensor reveals a
hidden similarity between the coupling to the 1D and 3D quantized radiation fields: $H_{3D,\mathrm{eff}}$ can be written
as weighted integrals of terms resembling $H_{\mathrm{eff}}$ with real-valued $k_{1D}\in[0,\, k_0]$
and imaginary-valued $k_{1D}\in[i0,+i\infty]$:
\begin{widetext}
\begin{equation}\label{Heff-3d}
H_{3D,\mathrm{eff}} = -i\frac{3\gamma_0}{4 k_0}\int_{0}^{k_0}\frac{d\tilde{k}}{2\pi} \rho_{+}(\tilde{k}) \sum_{m,n=1}^{N}e^{i\tilde{k}|z_m-z_n|}\;\sigma_m^\dagger \sigma_n
-\frac{3\gamma_0}{4k_0}\int_0^{+\infty}\frac{d\tilde{k}}{2\pi} \rho_{-}(\tilde{k}) \sum_{m,n=1}^{N}e^{-\tilde{k}|z_m-z_n|}\;\sigma_m^\dagger \sigma_n,
\end{equation}
\end{widetext}
where $\gamma_0$ is the spontaneous emission rate and $k_0$ is the resonant wavenumber. If the atoms are polarized
parallel to the chain, $\rho_{\pm}(\tilde{k})=2\pi(1\mp\tilde{k}^2/k_0^2)$ and the atom-atom interaction is short-range ($\sim 1/r^3$).
If the atoms are polarized transverse to the chain, $\rho_{\pm}(\tilde{k})=\pi(1\pm\tilde{k}^2/k_0^2)$ and the atom-atom interaction is
long-range ($\sim 1/r$).

In combination with the two exact features of our analytical results for $H_{\mathrm{eff}}$:
\begin{enumerate}
\item The leading order solutions of $\delta_{\xi}$ and $|\phi_{k_{\xi}}\rangle$ are independent of the values
of $k_{1D}$,see Eq. (\ref{ansatz});
\item The proportionality $\delta_{\xi}\propto\xi$ and the parabolic dispersion relation $\omega_k\propto\delta_{\xi}^2$ hold to order-$N^{{-}2}$,
for all values of  $k_{1D}$,
\end{enumerate}
this explains the universality of the $\xi^2/N^3$-scaling: Feature 1 implies that the
leading order eigenstates of $H_{\mathrm{eff}}$, shown in Eq. (\ref{ansatz}), are shared simultaneously by all terms integrated in $H_{3D, \mathrm{eff}}$,
and thus by the full $H_{3D,\mathrm{eff}}$ due to linearity.
Feature 2, hence implies that the corresponding decay rates, scaling as $\xi^2/N^3$, also apply to the subradiant states of $H_{3D,\mathrm{eff}}$.
The prerequisite is that $H^{I}_{3D,\mathrm{eff}}$ must have dark states with $k\approx\pm\pi/d$. Hence we require $k_0<\pi/d$ which implies that the ensemble is only sub-radiant in the 3D field if the atom-atom distance is less than half the resonant wave length \cite{Olmos:2013aa,Asenjo-Garcia2017}.

\emph{Subradiant multiply-excited states}-
When the number of atomic excitations $n_e\ll N$, the leading order Holstein-Primakoff (HP) approximation \cite{Holstein1940} usually applies
and one may replace
$\sigma^{\dagger}_m\sigma_n$ of Eq. (\ref{H-1d}) with the bosonic operators $b_m^{\dagger}b_n$ and obtain
a quadratic bosonic $H_{\mathrm{eff}}$. It works well for the superradiant modes with wavenumber $\pm k_{1D}$.
But for the subradiant multiply-excited states, the bosonic creation operators prepare
exchange symmetric combinations of subradiant one-excitation states with decay rates scaling
as $N^{-1}$ \cite{Asenjo-Garcia2017,Albrecht2018} which is much larger than the numerically observed $N^{-3}$-scaling
\cite{Asenjo-Garcia2017,Albrecht2018,Henriet:2019aa}.
Instead, the numerical results were found to favor fermionic exchange anti-symmetric combinations
of the subradiant one-excitation states \cite{Asenjo-Garcia2017,Albrecht2018,Henriet:2019aa}.

This somewhat surprising result inspires a closer scrutiny of the HP transformation. Including second order corrections due to saturation, the
HP transformation reads $\sigma_m=(1-b_m^\dagger b_m/2)b_m$ so that we can write
$H^{I}_{\text{eff}}=H_{\mathrm{SR}}+Q+Q^{\dagger}$ with the quadratic term
$H_{\mathrm{SR}}=N\Gamma_{1D}/4\sum_{\epsilon=\pm}b_{\epsilon k_{1D}}^\dagger b_{\epsilon k_{1D}}$ and quartic terms
\begin{equation}\label{HP transformation}
Q=-\frac{\Gamma_{1D}}{8}\sum_{\epsilon=\pm}\sum_{p,q} b_{\epsilon k_{1D}}^\dagger b^{\dagger}_{p+q-\epsilon k_{1D}} b_p b_q.
\end{equation}
Here, $b^{\dagger}_k=N^{-1/2}\sum_m e^{ikz_m}b^{\dagger}_m$ and the summation over
wavenumber is taken over an orthonormal basis $\{|k\rangle\}_{k}$ containing $|{\pm} k_{1D}\rangle$.
The quadratic term $H_{SR}$ has a prefactor $N$-times larger than those of $Q$ and $Q^{\dagger}$, but to assess their influence,
we should take account of not only the prefactors but also the magnitudes of
the operator terms. For $H_{SR}$, the magnitude of $b_{\epsilon k_{1D}}^\dagger b_{\epsilon k_{1D}}$ can be
estimated by its typical expectation value over the relevant Hilbert space, i.e., the subradiant states.

Reasonably, one may expect that a subradiant state contains no excitation of superradiant modes, i.e., typically
$\langle b_{\epsilon k_{1D}}^\dagger b_{\epsilon k_{1D}}\rangle\approx 0$. Thus the magnitude of $H_{SR}$ is suppressed.
Meanwhile, Eq. (\ref{HP transformation}) shows that
$Q$ annihilates a two-boson dark state, $b^{\dagger}_p b^{\dagger}_q|\varnothing\rangle$, with respect to $H_{SR}$, and generates a superradiant two-boson state
$b^{\dagger}_{\epsilon k_{1D}}b^{\dagger}_{p+q-\epsilon k_{1D}}|\varnothing\rangle$ ($|\varnothing\rangle$ denotes the boson vacuum).
This demonstrates that the saturation correction to the HP approximation plays a significant role even in the low excitation regime ($n_e\ll N$),
in contrast to its role in many other applications.

An effective theory for how $Q$ couples the dark states to superradiant states and hereby determines
their subradiant behavior is illustrated in the right panel of Fig. \ref{fig1}.
The effect of $Q$ and $Q^{\dagger}$ is distilled by eliminating
the superradiant states, in a manner similar to the adiabatic elimination of excited
state manifolds to restrict the effective dynamics of quantum systems to their ground state manifold \cite{Reiter2012}.

Note that the subset of superradiant states with only a single excitation of
the superradiant modes and thus the eigenvalue (decay rate) $N\Gamma_{1D}/4$, has the strongest coupling to the dark/subradiant states.
We hence disregard the coupling to other superradiant states and the effective coupling among subradiant states reduces to
$V_{\mathrm{sub}}=\frac{4}{N\Gamma_{1D}}P_{DS} Q^\dagger P_{SRS} Q P_{DS}$, with projection operators $P_{DS(SRS)}$ on the dark and superradiant spaces, respectively.
To evaluate this expression we use the operator relation that
$(b_{p'+q'-\epsilon' k_{1D}}b_{\epsilon' k_{1D}})(b^\dagger_{\epsilon k_{1D}}b^\dagger_{p+q-\epsilon k_{1D}})=\delta_{\epsilon,\epsilon'}\delta_{p'+q',p+q}$,
i.e., no population of the superradiant two-boson modes within the dark/subradiant space.
Finally, we obtain
\begin{equation}\label{vv}
\begin{aligned}
V_{\mathrm{sub}} & = \frac{1}{8N}\Gamma_{1D}\sum_{p,q,k} b^\dagger_{-p+q+k}b^\dagger_{p} b_{q} b_{k} \\
&=\frac{1}{8}\Gamma_{1D}\sum_{m=1}^{N}(b_m^{\dagger})^2(b_m)^2.
\end{aligned}
\end{equation}
That is, $V_{\mathrm{sub}}$ induces decay with rate $O(\Gamma_{1D})$ of nominally subradiant
states having more than a single HP boson excitation at the same site.

In the absence of $V_{\mathrm{sub}}$, approximate multiply-excited states are created by the operators
$b^{\dagger}_{\xi}=\sum_{m}\langle e_m|\phi_{k_{\xi}} \rangle b^{\dagger}_{m}$, with $|\phi_{k_{\xi}} \rangle$ the one-excitation eigenstates
Eq. (\ref{ansatz}) of $H_{\mathrm{eff}}$. As $V_{sub}$ cannot be treated as a perturbation,
we study the effective Hamiltonian $\mathcal{H}=\frac{1}{2}\sum_{\xi}\gamma_{\xi}b_{\xi}^\dagger b_{\xi}+V_{\mathrm{sub}}$,
where only the most subradiant states ($\xi\ll N$) are included in the sum. In the Supplemental Material we show that
in the continuous limit, $\mathcal{H}$ can be written as
the second-quantized form of the Hamiltonian
\begin{equation}\label{Tonks-Girardeau, Lieb-Liniger}
\mathcal{H}=
\sum_{i=1}^{n_e}\bigg[\frac{-\partial_{x_i}^2}{2m_{*}}+ V(x_i)\bigg]
+ 2c_{LL}\sum_{i< j=1}^{n_e}\delta(x_i-x_j),
\end{equation}
where $c_{LL}=d\Gamma_{1D}/8$ and $V(x_i)$ is a 1D box potential for bosons in the interval $[0, Nd]$ with one-excitation eigenstates $|\phi_{k_{\xi}}\rangle$
given by Eq. (\ref{ansatz}).
The observation behind Eq. (\ref{Tonks-Girardeau, Lieb-Liniger}) is that the $\xi^2/N^3$-scaling of $\gamma_{\xi}$ takes the same form of a kinetic energy
$\gamma_{\xi}=k_{\xi}^2/(2m_{*})$ when $k\approx 0$; or the kinetic energy in a gauge field
$\gamma_{\xi}=(k_{\xi}+\pi/d)^2/(2m_{*})$ when $k\approx -\pi/d$.  With the parametrization of the model,
the effective mass in the kinetic energy term reads $m_{*}=\xi^2\pi^2/(N^2 d^2\gamma_{\xi})\propto N$.

\begin{figure}[b]
  \centering
    \includegraphics[width=0.95\textwidth]{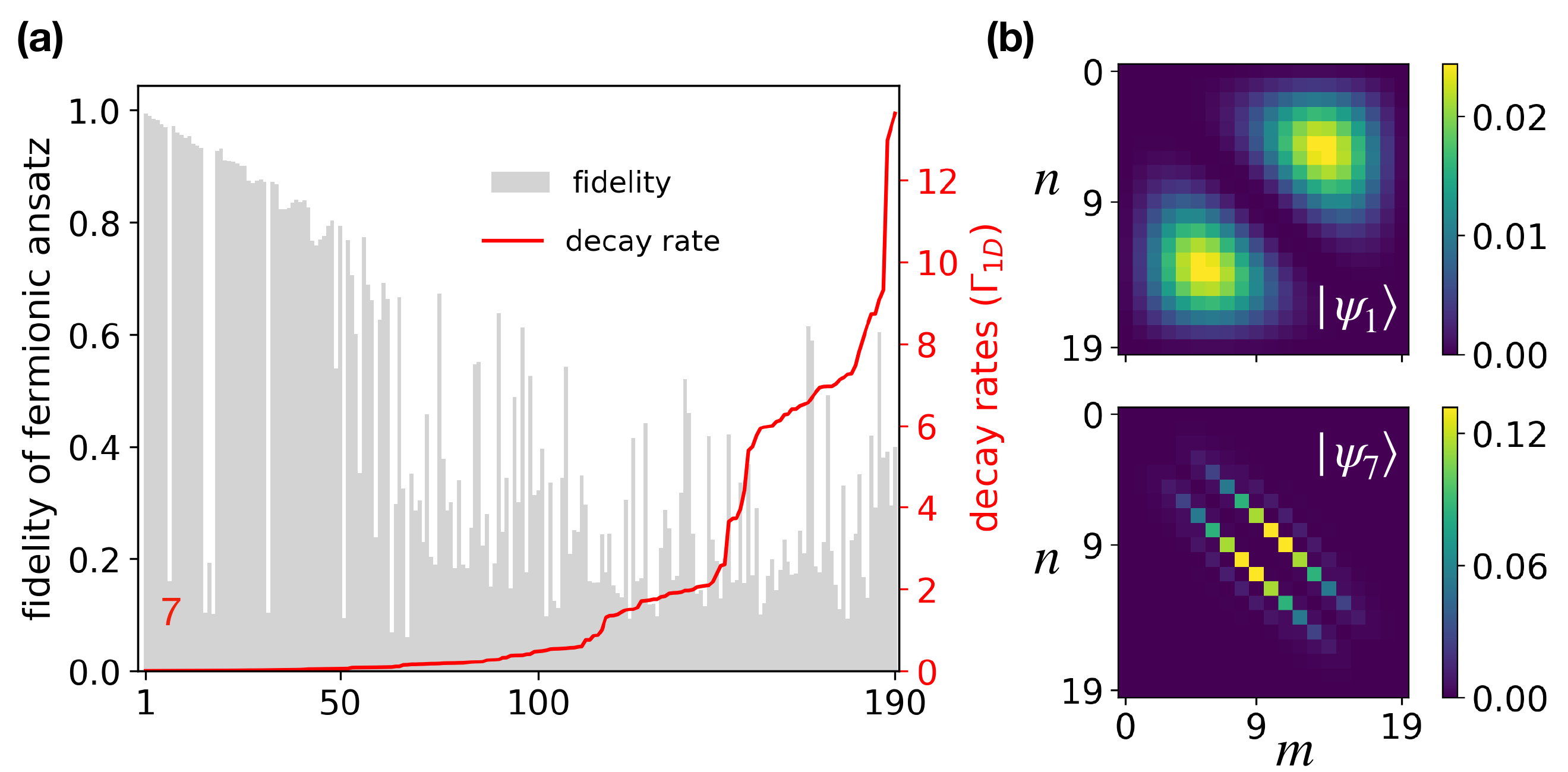}
\caption{
(a) The two-excitation eigenstates of system with
$k_{1D}=0.2\pi/d$ and $N=20$ are sorted by increasing decay rates. The bars show the maximal fidelity that a fermionic ansatz can achieve for each eigenstate.
The fermionic ansatz fits a broad range of the most subradiant states while
a few exceptional states (the dips in the fidelity, e.g., state number 7) show distinct non-fermionic behavior.
(b) Position distributions of the atomic excitations, $|\langle\psi_{1(7)}|e_m, e_n\rangle|^2$ with $m(n)=0,1,\cdots 19$,
of a typical fermionic subradiant state, $|\psi_1\rangle$ (upper panel) and the non-fermionic state $|\psi_7\rangle$ (lower panel).
The lower panel feature at $|z_m-z_n |\approx 2d$ indicates that $|\psi_7\rangle$ supports a dimer-like bound excitation.
}

\label{fig2}
\end{figure}

We recognize Equation (\ref{Tonks-Girardeau, Lieb-Liniger}) as  the Lieb-Liniger model \cite{Lieb1963}
originally proposed for 1D gases of hard-core bosons.
As the effective mass $m_{*}$ diverges in the large $N$ limit, the kinetic energy-like part of Eq. (\ref{Tonks-Girardeau, Lieb-Liniger}) vanishes.
This implies that
Eq.(\ref{Tonks-Girardeau, Lieb-Liniger}) reaches the Tonks-Girardeau regime \cite{Tonks1936,Girardeau1960} of the Lieb-Liniger model,
where the eigenstates of $\mathcal{H}$ can be obtained
via a fermion-boson mapping \cite{Girardeau1960,Cazalilla2011}:
For a free fermion model described by $\sum_{\xi}\frac{1}{2}\gamma_{\xi}f_{\xi}^{\dagger}f_{\xi}$,
where $f^{\dagger}_{\xi}=\sum_{m} \langle e_m|\phi_\xi\rangle f^{\dagger}_m$, we write down
its eigenstates (e.g., two-fermion states) $f^{\dagger}_{\xi_1}f^{\dagger}_{\xi_2}|\varnothing\rangle$,
and replace $f^{\dagger}_{m} f^{\dagger}_{n}$ with
$\mathrm{sign}(n-m)b^{\dagger}_{m} b^{\dagger}_{n}$, where $\mathrm{sign}(n-m)$ is necessary to ensure
the consistency with the fermionic commutation relation. This yields a fermion-like bosonic state
$|F_{\xi_1, \xi_2}\rangle=\sum_{m<n}[\phi_{k_{\xi_1}}(z_m)\phi_{k_{\xi_2}}(z_n)
-e^{i\varphi}\phi_{k_{\xi_2}}(z_m)\phi_{k_{\xi_1}}(z_n)]b^{\dagger}_{m} b^{\dagger}_{n}|\varnothing\rangle$
($e^{i\varphi}=1$ is introduced for later convenience). Since $b^{\dagger}_{m} b^{\dagger}_{n}|\varnothing\rangle
= |e_m,e_n\rangle$, we recover the fermionic ansatz of the two-excitation sector \cite{Asenjo-Garcia2017,Albrecht2018,Henriet:2019aa}.
As a direct consequence of their representation as non-interacting fermions the
decay rates of the most subradiant multiply-excited states are merely the sum of the decay
rates of their one-excitation constituents, e.g., $\gamma_{\xi_1}+\gamma_{\xi_2}$.
This explains the numerical observations of Refs. \cite{Asenjo-Garcia2017,Albrecht2018,Henriet:2019aa}.
The above mapping also applies to more excitations \cite{Girardeau1960,Cazalilla2011}.

For a finite atom chain, $m_{*}$ is finite and Eq. (\ref{Tonks-Girardeau, Lieb-Liniger}) deviates from
the Tonks-Girardeau limit, hence the fermionic ansatz, for example the
two-excitation state $|F_{\xi_1, \xi_2}\rangle$, deviates slightly from
the numerical eigenstates denoted by $|\psi^{\mathrm{num}}_{\xi_1, \xi_2}\rangle$. The deviation
quantified by the infidelity $1-|\langle F_{\xi_1, \xi_2}|\psi^{\mathrm{num}}_{\xi_1, \xi_2}\rangle|^2$, is numerically found to
scale as $N^{-2}$ when both components, $|\phi_{k_{\xi_1}}\rangle$ and $|\phi_{k_{\xi_2}}\rangle$,
come from the same branch of the one-excitation subradiant states,
i.e., $k_{\xi_1},\,k_{\xi_2}\approx\pm\pi/d$ (or both $\approx 0$); otherwise,
the infidelity scales as $N^{-1}$ (when $k_{\xi_1}\approx \pm \pi/d$ and $k_{\xi_2}\approx 0$) \cite{Albrecht2018}.
These behaviors can also be explained from
the Lieb-Liniger model of Eq. (\ref{Tonks-Girardeau, Lieb-Liniger}). The fermion-boson mapping is not exact
and the phase factor $e^{i\varphi}$ introduced above deviates from unity by a factor in the form of
$(k_{\xi_1}-k_{\xi_2})/(m_{*}c_{LL})$ \cite{Lieb1963}. Since $k_{\xi_1}-k_{\xi_2}$ is $O(N^{-1})$ or $\approx\pi/d$ in the two cases considered,
while $m_{*}c_{LL}$ scales as $N$, their ratio scales exactly in the same manner as the numerically observed
infidelities \cite{Albrecht2018}. Larger discrepancies with the fermionic ansatz are detectable when the decay rates increase.

\emph{Universality-}The mapping to Lieb-Liniger model and the Tonks-Girardeau gas can also be extended to
the 1D atomic chain coupled to 3D free-space modes described by Eq. (\ref{Heff-3d}). Here
$H^{I}_{3D,\mathrm{eff}}$ possesses short-lived eigenstates $|\tilde{k}\rangle$
with $\tilde{k}\in[{-}k_0, k_0]$ and different decay rates $\gamma_{\tilde{k}}$. Each of them
will contribute to $V_{\mathrm{sub}}$ a term with prefactor $\gamma_{\tilde{k}}/N^2$. Hence we have
$V_{\mathrm{sub}}\propto \sum_{\tilde{k}}\gamma_{\tilde{k}}/N^2= \gamma_0/N$, similar to the coefficient in the first line of Eq. (\ref{vv}).
Since the $\xi^2/N^3$-scaling decay rates apply in the one-excitation sector in 3D free-space, the
fermionic ansatz also applies here. Since only subradiant states with $k\approx\pm\pi/d$ appear in 3D free-space,
the pertaining $N^{-2}$ scaling applies to the infidelities
of all states given by the fermionic ansatz. This matches the numerical results \cite{Asenjo-Garcia2017}.

\emph{ Conclusion and Discussion-}
In this Letter, we have developed a theory to explain the $\xi^2/N^3$-scaling of subradiant decay rates
and the fermionic behavior of multiply-excited subradiant states identified in numerical calculations on 1D
atom chains coupled to both 1D and 3D radiation reservoirs \cite{Asenjo-Garcia2017,Albrecht2018,Henriet:2019aa}.
We find that the universal $\xi^2/N^3$-scaling results from a parabolic dispersion relation of the atomic
excitation, and imaginary corrections of the Bloch quasi-momentum eigenstates of the non-Hermitian Hamiltonian.
For multiply excited systems quartic corrections to the Holstein-Primakoff (HP) expansion of the effective spin
Hamiltonian for the atom chain dominate the coupling of the sub-radiant states, and lead to a formulation equivalent
to the Lieb-Liniger model of a 1D bosonic quantum gas in the Tonks-Girardeau regime \cite{Tonks1936,Girardeau1960}.
The fermionic ansatz solution of that problem explains the decay rates and the properties of the solutions
found in Ref. \cite{Asenjo-Garcia2017,Albrecht2018}.
There is a high current interest and many potential applications
of subradiance \cite{Chang2012,Gonzalez-Tudela2017,
Plankensteiner2017,Paulisch2016} and the analytical findings presented here may inspire further study of subradiance in light-matter
interactions of more complex geometries, e.g., systems with higher
dimensional atom arrays \cite{Facchinetti:2016aa}, chiral waveguides that break the parity symmetry \cite{Ramos2014} and
setups with topological effects \cite{Perczel2017,Ozawa2018}.

Let us conclude by discussing a remaining theoretical issue. We recall
our effective separation of the Hamiltonian into an interaction term, $V_{\mathrm{sub}}$, based on $H^{I}_{\mathrm{eff}}$
and an expansion on subradiant eigenmodes $b_{\xi}$ for which $H^{R}_{\mathrm{eff}}$ contributes the decay rates $\gamma_{\xi}$.
Like the numerical calculations, a more rigorous analytical approach should incorporate $H^{R}_{\mathrm{eff}}$ and $H^{I}_{\mathrm{eff}}$
on an equal footing. The fact that our separate treatment applies may be understood from the perturbation view. The leading order
approximation of the subradiant states are the dark states of $H^{I}_{\mathrm{eff}}$. They are also approximate eigenstates of
$H^{R}_{\mathrm{eff}}$ when restricted to the most subradiant states. It means that the fermionic ansatz,
as the leading order approximation, is shared by both
$H^{I}_{\mathrm{eff}}$ and $H_{\mathrm{eff}}$. Therefore analyzing $V_{\mathrm{sub}}$ from the simpler $H^{I}_{\mathrm{eff}}$ is
sufficient to capture the salient fermionic behavior.
This is also verified by a direct construction of the fermionic ansatz without using the HP transformation,
for both $H^{I}_{\mathrm{eff}}$ and $H_{\mathrm{eff}}$ (see the Supplemental Material).
Interestingly, we find that the fermionic ansatz does not exhaust all the most subradiant eigenstates.
For a medium-size ensemble of $N=20$ atoms, we obtained numerical eigenstates of $H_{\mathrm{eff}}$ with very low fermionic state fidelity.
The subradiant states of this different character have well defined ``center of mass'' wavenumber, and well
defined spatial separation, as illustrated in Fig. \ref{fig2}(b). Further discussion of these states is beyond the scope of this manuscript,
but may be of interest for future work possibly together with the interesting prospects for studying quantum fluctuations \cite{Paredes2004,Jacqmin2011,Budde2004} in the Tonk-Girardeau gas theory by detection of the excited state correlations among atoms in a subradiant chain.

\emph{Acknowledgments}-This work was supported by the Villum Foundation and by the European Unions Horizon 2020 research and
innovation program (Grant No. 712721, NanOQTech).

%


\section{Supplemental Material}

\subsection{A. Energy Levels of the Subradiant States}
We derived the decay rates of the subradiant states in the one-excitation sector. As a byproduct, their
energy levels are given as following: for the subradiant states with $k_{\xi}\approx \xi\pi/(Nd)\;(\xi\ll N)$, we have
\begin{equation}
\omega_\xi=\frac{\Gamma_{1D}}{2}\cot(\frac{k_{1D}}{2}d)+\Gamma_{1D}\frac{\cos(k_{1D} d/2)}{\sin^{3}(k_{1D} d/2)}(\frac{\xi\pi}{2N})^2;
\end{equation}
for those with $k_{\xi}\approx -\pi/d+\xi\pi/(Nd)$ ($\xi\ll N$), we have
\begin{equation}
\omega_\xi=-\frac{\Gamma_{1D}}{2}\tan(\frac{k_{1D}}{2}d)-\Gamma_{1D}\frac{\sin(k_{1D} d/2)}{\cos^{3}(k_{1D} d/2)}(\frac{\xi\pi}{2N})^2.
\end{equation}
Apart from the constant part, the above expressions show that the subradiant states have Lamb shifts
proportional to $\xi^2/N^2$. The band is parabolic and flat around the extreme point $k=0$ or $k=\pm\pi/d$.

\subsection{B. Hamiltonian of 1D Atom Chain in 3D Free-Space}
The effective atom-atom coupling Hamiltonian is expressed as
\begin{equation}
H_{3D,\mathrm{eff}}=-\mu_0 \omega_0^2\sum_{i,j=1}^{N}
\mathbf{d}^*_i\cdot\mathbf{G}(\mathbf{r}_i, \mathbf{r}_j, \omega_0)\cdot\mathbf{d}_j
\;\sigma_i^{\dagger}\sigma_j,
\end{equation}
where $\mathbf{d}_i$ and $\mathbf{r}_i$ is the dipole moment and the position of the \emph{i}th atom, respectively, and
$\mu_0$ is the vacuum permeability. For the case of 3D free-space, the dyadic Green's tensor $\mathbf{G}(\mathbf{r}_i, \mathbf{r}_j,\omega)=
\mathbf{G}(\mathbf{r}_i-\mathbf{r}_j,\omega)$ is given as
\begin{equation}
\begin{aligned}
\mathbf{G}(\mathbf{r}, \omega_0)=\frac{e^{ik_0r}}{4\pi k_0^2r^3}\bigg[ (k_0^2 r^2 & +ik_0r-1)\mathbf{I}+ \\
& (-k_0^2 r^2-3ik_0 r+3)\frac{\mathbf{r}\mathbf{r}}{r^3}\bigg]
\end{aligned}
\end{equation}
where $k_0=\omega_0/c$. This expression can be transformed to the wave number presentation and
yields $H_{3D,\mathrm{eff}}$ presented as Eq. (6) of the main text.

\subsection{C. Transformation to Continuous Limit}
Equation (9) of the main text is written in a discrete notation. The continuous expression can be obtained from the discrete notation by the mapping
\begin{equation}
\sum_{i=1}^{N}\rightarrow \frac{1}{d}\int_{0}^{Nd} dx,\qquad b_{i}\rightarrow \sqrt{d}\, b_{x}.
\end{equation}
The bosonic commutation relation changes from $[b_i, b_j^{\dagger}]=\delta_{i,j}$ to $[b_x, b_y^\dagger]=\delta(x-y)$.

\subsection{D. Direct construction}
To see why the fermionic ansatz has the $N^{-3}$ decay rate, we introduce the two-excitation
state $|k_1,k_2\rangle=\sum_{m<n} e^{ik_1 z_m+ik_2 z_n} |e_m, e_n\rangle$,
and evaluate
\begin{equation}\label{imag2e}
\begin{aligned}
H^{I}_{\text{eff}}\,|k_1, k_2\rangle  =  \frac{N\Gamma_{1D}}{4} & \sum_{\epsilon=\pm}
\bigg[  g_{k_1, \epsilon} |b_{k_2, \epsilon k_{1D}}\rangle\\
- h_{k_2, \epsilon} |b_{k_1, \epsilon k_{1D}}\rangle
+ & c_{k_1,k_2,\epsilon} |b_{k_1+k_2-\epsilon k_{1D}, \epsilon k_{1D}}\rangle\bigg],
\end{aligned}
\end{equation}
where
\begin{equation}
\begin{aligned}
g_{k,\epsilon} & =\frac{1}{N}\frac{1}{e^{-i(k-\epsilon k_{1D})d}-1}, \\
h_{k, \epsilon} & =\frac{1}{N} \frac{e^{i(k-\epsilon k_{1D})Nd}}{e^{-i(k-\epsilon k_{1D})d}-1},
\end{aligned}
\end{equation}
and
$c_{k_1, k_2, \epsilon}{=}g_{k_1,\epsilon}^* + g_{k_2,\epsilon}$ and $|b_{k,k'}\rangle{=}|k,k'\rangle +|k',k\rangle$.
All these state amplitudes scale as $N^{-1}$ rather than the desired $N^{-2}$-scaling, which is required to obtain the $1/N^3$-scaling of the decay rates.

As in the one-excitation sector, we may proceed and construct superpositions of four degenerate states
$|k_1, k_2\rangle$, $|{-}k_1, k_2\rangle$, $|k_1,{-}k_2\rangle$ and $|{-}k_1, {-}k_2\rangle$, to
$|\phi_{k_1}, \phi_{k_2}\rangle=\sum_{m<n}\phi_{k_1}(z_m)\phi_{k_2}(z_n)|e_m, e_n\rangle$,
where $\phi_{k}(z_m)=\langle e_m|\phi_k\rangle$. Then in the expression of $H^{I}_{\text{eff}}|\phi_{k_1}, \phi_{k_2}\rangle$,
the state amplitudes on $|b_{k_{1(2),\epsilon k_{1D}}}\rangle$, but
not on $|b_{k_1+k_2-\epsilon k_{1D}, \epsilon k_{1D}}\rangle$, are successfully reduced to the $N^{-2}$-scaling.

To also suppress the latter,
we form the superposition with the permuted state $|\phi_{k_2}, \phi_{k_1}\rangle$. The suitable superposition turns out to be ``fermionic'', i.e.,
$|F_{k_1,k_2}\rangle\propto |\phi_{k_1}, \phi_{k_2}\rangle-|\phi_{k_2}, \phi_{k_1}\rangle$.
Different from what we have seen in the one-excitation sector, here the ``tails'' cannot be erased completely by the superposition.
It means that the Fermionic ansatz is only the leading order solution.

With reference to our concluding remarks in the main text on the full treatment of $H_{\mathrm{eff}}^R$ and $H_{\mathrm{eff}}^I$, the same construction can be applied for the full Hamiltonian $H_{\mathrm{eff}}$, where we obtain
\begin{equation}
\begin{aligned}\label{heffk1k2}
H_{\mathrm{eff}}|k_1, k_2\rangle=& (\omega_{k_1}+\omega_{k_2})|k_1, k_2\rangle \\
 - i \frac{N\Gamma_D}{2} & \bigg(
g_{k_1,+}|b_{k_D, k_2}\rangle-h_{k_2,-}|b_{-k_D, k_1}\rangle \\
& \quad +c_{k_1, k_2, -}|-k_D, k_1+k_2+k_D\rangle \\
&\qquad +c_{k_1, k_2,+}|k_1+k_2-k_D, k_D\rangle
\bigg).
\end{aligned}
\end{equation}
While Eq. (\ref{heffk1k2}) features  ``tails'' that are neither symmetric nor anti-symmetric, its main features, and hence the applicability of the fermionic ansatz,  are captured by $H^{I}_{\mathrm{eff}}$ given in Eq. (\ref{imag2e}).
As we show in the main text, the ``tails'' $|b_{k_1+k_2-\epsilon k_{1D}, \epsilon k_{1D}}\rangle$ of Eq. (\ref{imag2e}) arise via the second order terms in the Holstein-Primakoff (HP) transformation, which in turn establish the Tonks-Girardeau limit of the Lieb-Liniger model, and hence the fermionic solutions.

\end{document}